\documentstyle[11pt,newpasp,twoside,epsf]{article}
\def\edcomment#1{\iffalse\marginpar{\raggedright\sl#1\/}\else\relax\fi}
\reversemarginpar

\begin{document}
\title{The superbubble model for LiBeB production and Galactic
evolution}
\author{Etienne Parizot and Luke Drury}
\affil{Dublin Institute for Advanced Studies, 5 Merrion Square, Dublin
2, Ireland}

\begin{abstract}
We show that the available constraints relating to $^{6}$LiBeB
Galactic evolution can be accounted for by the so-called superbubble
model, according to which particles are efficiently accelerated inside
superbubbles out of a mixture of supernova ejecta and ambient
interstellar medium.  The corresponding energy spectrum is required to
be flat at low energy (in $E^{-1}$ below 500~MeV/n, say), as expected
from Bykov's acceleration mechanism.  The only free parameter is also
found to have the value expected from standard SB dynamical evolution
models.  Our model predicts a slope 1 (primary) and a slope 2
(secondary) behaviour at respectively low and high metallicity, with
all intermediate slopes achieved in the transition region, between
$10^{-2}$ and $10^{-1}Z_{\odot}$.
\end{abstract}

\section{Introduction}

Galactic nucleosynthesis and chemical evolution are about how a given
element is produced in the universe and how its abundance evolved from
the primordial universe on.  In the case of the light elements, it is
widely agreed that the nucleosynthesis occurs through spallation
reactions induced by energetic particles (EPs) interacting with the
interstellar medium (ISM).  In these reactions, an heavier nucleus
(most significantly C, N or O) is `broken into pieces' and transmuted
into one of the lighter $^{6}$Li, $^{7}$Li, $^{9}$Be, $^{10}$B or
$^{11}$B nuclei.  Except for $^{7}$Li, this spallative nucleosynthesis
is thought to be the main (if not the only) light element production
mechanism.  The case of $^{11}$B is slightly more complicated, as
neutrino-induced spallation in supernovae (the so-called
$\nu$-process) is sometimes invoked to increase the B/Be and
$^{11}\mathrm{B}/^{10}\mathrm{B}$ ratios which one would expect should
the light elements be produced by nucleo-spallation alone.

Concerning the Galactic evolution of light element abundances, Fields
et al.  (2000) have recently re-analyzed the available data as a
function of O/H, discussing the uncertainties associated with the
methods used to derive the O abundance, the stellar parameters and the
incompleteness of the samples.  According to their results, Be and B
evolution can be described in terms of two distinct production
processes: i) a primary process dominating at low metallicity and
leading to a linear increase of the Be and B abundances with respect
to O -- `slope 1' -- followed by ii) a secondary process compatible
with the standard expectations of the Galactic cosmic ray
nucleosynthesis scenario (GCRN) -- `slope 2'.  This behaviour is
characterized by a transition metallicity,
$Z_{\mathrm{t}}\equiv(\mathrm{O/H})_{\mathrm{t}}$, below which the
Be/O and B/O ratios are constant and above which they are proportional
to O/H. Although the value of $Z_{\mathrm{t}}$ is rather uncertain
because very few data points have yet been reported at $Z <
Z_{\mathrm{t}}$, energetics arguments show that a primary process
\emph{is} indeed required below, say, $10^{-2}Z_{\odot}$ (Parizot \&
Drury 1999a,b,2000b, Ramaty et al.  2000), and therefore the very
existence of a transition metallicity separating a primary from a
secondary evolution scheme seems reasonably well established.  In
spite of the current large uncertainties on the exact value of
$Z_{\mathrm{t}}$, Fields et al.  find a range of possible values
between $10^{-1.9}$ and $10^{-1.4}(\mathrm{O/H})_{\odot}$ (see also
Olive, this conference).

This two-slope picture seems to reconcile the two competing theories
for light element nucleosynthesis, namely GCRN which predicts a
secondary behaviour for Be and B evolution (e.g. Vangioni-Flam et al. 
1990, Fields \& Olive 1999), and the superbubble model which predicts
a primary behaviour at low metallicity (Parizot \& Drury 1999b,2000b,
Ramaty et al.  2000).  However, we show here that when applied to the
whole lifetime of the Galaxy (not only to the early Galaxy), the
superbubble model \emph{alone} actually predicts the entire two-slope
behaviour inferred from observations, and accounts for all the
qualitative and quantitative constraints currently available. 
Implications for particle acceleration inside a superbubble (SB) are
also analyzed.

\section{Description of the SB model}

The superbubble model is based on the observation that most massive
stars are born in associations, and evolve quickly enough to explode
as SNe in the vicinity of their parent molecular cloud.  The dynamical
effect of repeated SN explosions in a small region of the Galaxy is to
blow large bubbles -- superbubbles -- of hot, rarefied material,
surrounded by shells of swept-up and compressed ISM. The interior of
superbubbles consists of the ejecta and stellar winds of evolved
massive stars \emph{plus} a given amount of ambient ISM evaporated off
the shell and dense clumps passing through the bubble.  The exact
fraction of the ejecta material inside the SB is not well known, and
can be expected to vary with time and from one SB to another. 
However, this fraction, which we note $x$, is all we need to know in
order to fully determine the mean composition of the matter inside
SBs.  Noting $\alpha_{\mathrm{ej}}(\mathrm{X})$ and
$\alpha_{\mathrm{ISM}}(\mathrm{X})$ the abundances of element X among
the SN ejecta and in the ISM, respectively, we can indeed write the
abundance of X inside the SB as:
\begin{equation}
	\alpha_{\mathrm{SB}}(\mathrm{X}) =
	x\alpha_{\mathrm{ej}}(\mathrm{X})
	+ (1 - x)\alpha_{\mathrm{ISM}}(\mathrm{X}).
	\label{SBAbundances}
\end{equation}

The second assumption of the SB model is that the material inside
superbubbles is efficiently accelerated by a combination of shocks
produced by SN explosions and supersonic stellar winds, secondary
shocks reflected by other shocks or clumps of denser material, and a
strong magnetic turbulence created by the global activity of all the
massive stars.  Two different SB models have been proposed so far,
assuming different EP compositions and energy spectra.  In our model
(Parizot \& Drury 1999b,2000b), we follow Bykov \& Fleishman (1992) and
Bykov (1995,1999) and argue that the SB acceleration process produces
a rather flat spectrum at low energy, namely in $E^{-1}$, as expected
from multiple shock acceleration theory (Markowith \& Kirk 1999), up
to a few hundreds of MeV/n, say.  Above this value, the spectrum of
the superbubble EPs (SBEPs) is either cut off through a steep
power-law or turned into the standard cosmic ray source spectrum
(CRS), in $E^{-2}$.  The exact behaviour of this so-called `SB
spectrum' at high energy is important in itself and should be derived
from a detailed calculation of the particle acceleration, but we do no
consider it here, as it is not relevant to our problem (most of the
LiBeB production arises from the most numerous low-energy particles
anyway).  The other SB model proposed so far (Ramaty \& Lingenfelter
1999, Ramaty et al.  2000) assumes that the SBEPs \emph{are} actually
the cosmic rays and thus their energy spectrum is the standard CRS
spectrum ($Q(p)\propto p^{-2}$).

To summarize, the essence of the SB model is that repeated SN
explosions occurring in OB associations lead to the acceleration of EPs
having either the CRS spectrum or the SB spectrum, and a composition
given by Eq.~(1), where the only free parameter is the proportion of
the ejecta inside the SB: $x$.  In principle, $x$ can be derived from
the study of SB evolution dynamics, coupled with a gas evaporation
model.  But we shall first study LiBeB evolution for itself with no
external prejudice about the value of $x$, and therefore consider it
as a free parameter which we vary from 0 (i.e. SBEPs have the ambient
ISM composition) to 1 (i.e. SBEPs are made of pure SN ejecta).  Later,
we compare the value derived from the LiBeB constraints with the value
expected from standard SB dynamical models.

\section{Be and B Galactic evolution}

\subsection{Qualitative features}

Having parameterized our problem as above, we can easily calculate the
Be/O production ratio in the Galaxy as a function of
$Z_{\mathrm{ISM}}\equiv (\mathrm{O/H})_{\mathrm{ISM}}$.  We consider
SBs blown by 100 SNe exploding continuously over a lifetime of 30~Myr. 
We then integrate the Be production rates induced by the SBEPs and
divide the result by the total O yield (added up assuming a Salpeter
IMF and SN yields from Woosley \& Weaver 1995).  The result is plotted
in Fig.~1 for various values of $x$ and the two investigated spectra. 
The main difference between the latter is the Be production
efficiency, i.e. the number of Be produced per erg of SBEP. Apart from
the SB spectrum being more efficient, both figures show distinctively
the sought-for two-slope behaviour, with a transition metallicity
$Z_{\mathrm{t}}$ depending on the actual value of $x$.  This behaviour
derives directly from Eq.~(1).  Replacing X by O there, we see that
the abundance of O among the SBEPs is essentially
$x\alpha_{\mathrm{ej}}(\mathrm{O})$ at low metallicity, and
$(1-x)\alpha_{\mathrm{ISM}}(\mathrm{O})$ above a transition
metallicity $Z_{\mathrm{t}} \sim \frac{x}{1-x}Z_{\mathrm{ej}}$ (where
$Z_{\mathrm{ej}}\sim 10 Z_{\odot}$).  Therefore, remembering that O is
the main progenitor of Be, we find that the SB model predicts a
primary behaviour below $Z_{\mathrm{t}}$ (production efficiency
independent of $Z_{\mathrm{ISM}}$), and a secondary behaviour above
$Z_{\mathrm{t}}$, since the SB model is then essentially identical to
the GCRN model (except maybe for the assumed energy spectrum).

Incidentally, it is interesting to note that the SB model can be
considered as a correction of the GCRN scenario, taking into account
the chemical inhomogeneity of the early Galaxy.  Indeed, since
particle acceleration occurs precisely in those places where metals
are released (i.e. superbubbles), the SBEP composition is considerably
richer in O than the average ISM, as long as the SN ejecta dominate
the O content of the SBs.  Afterwards, it makes little difference, as
far as composition is concerned, whether the EPs producing LiBeB are
accelerated inside SBs or in the regular ISM.

\begin{figure}
\plottwo{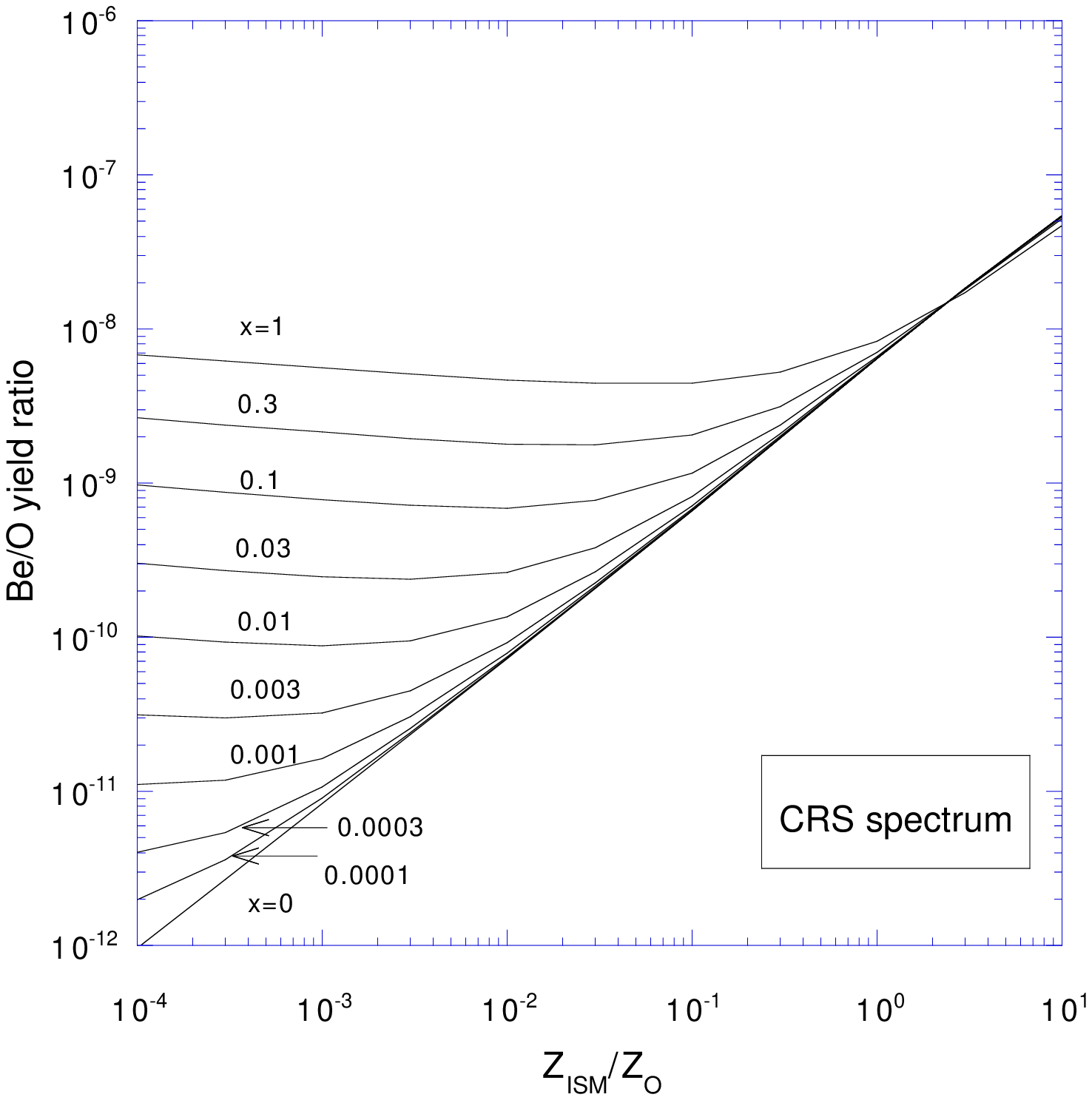}{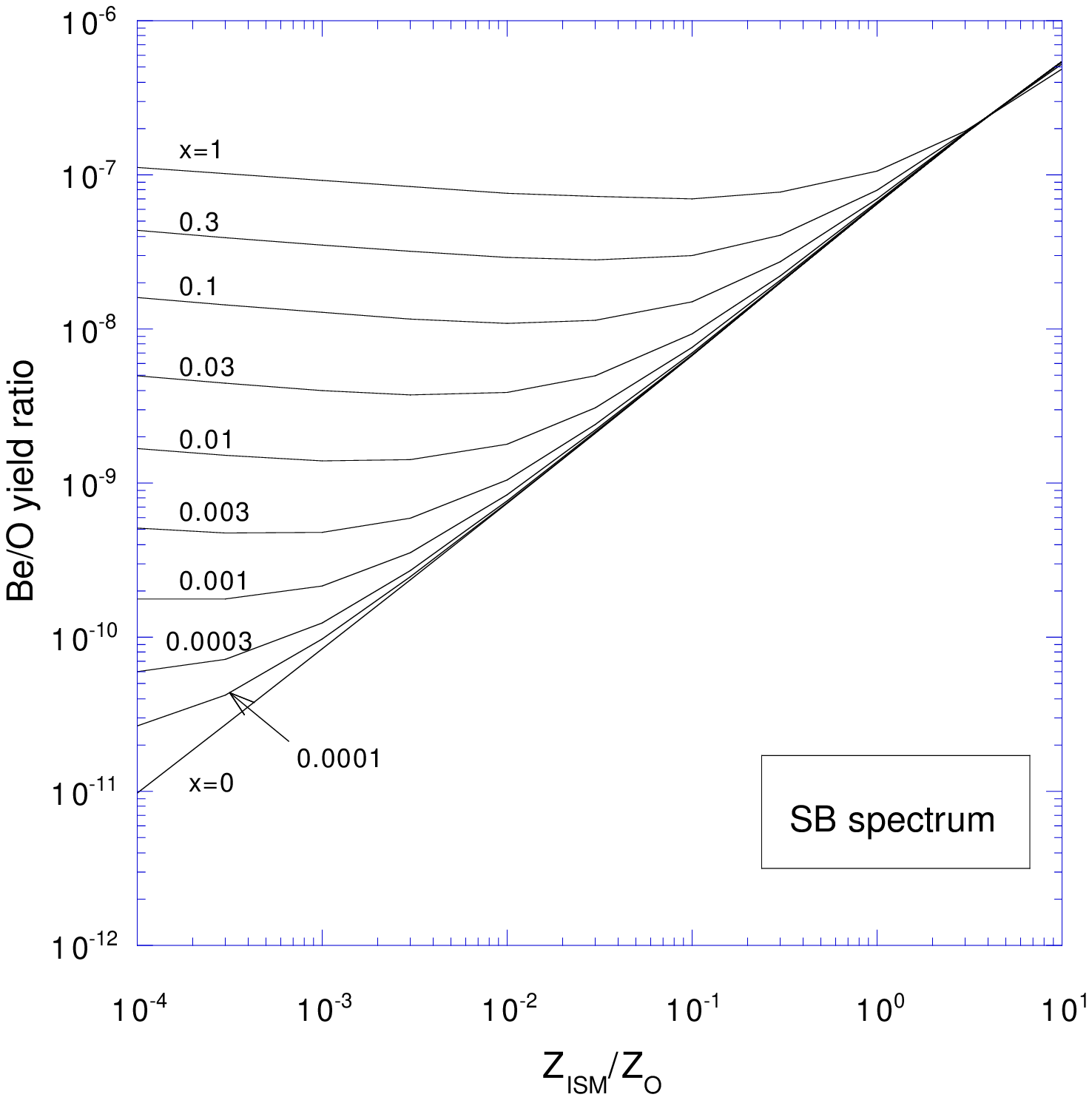}
\caption{Be/O yield ratios obtained with the CRS spectrum (left) and
the SB spectrum (right), as a function of the ambient metallicity, for
various values of the mixing parameter, $x$.  The Be yield is
calculated for a SBEP total energy of $10^{50}$ erg per SN.}
\end{figure}

Finally, we see from Fig.~1 that the predicted slope 1 and slope 2
correlations between Be and O are limit behaviours for very low and
very high metallicity respectively.  Depending on the value of $x$,
any intermediate value for the Be-O slope is reached over a given
range of stellar metallicity.  This is in contrast with what would
arise if the two-slope behaviour were to be explained in terms of two
different mechanisms (e.g. the SB model at low $Z$ and GCRN at high
$Z$).  In that case, indeed, one would have a sharp change of slope at
the precise metallicity where the secondary process becomes dominant,
with no intermediate values.  Of course, expected physical
fluctuations of the parameters would weaken this effect, and current
observational error bars prevent us from distinguishing conclusively
between the two pictures.  But we argue that the observed `slope 1.45'
behaviour reported by Boesgaard \& Ryan (this conference) can be
explained (in principle) only if there is a continuous
\emph{transition} from slope 1 to slope 2 within a \emph{unique} model
(as in the SB model above ), rather than two unrelated models with a
slope 2 eventually superseding a slope 1.

\subsection{Quantitative features}

Quantitatively, the Be/O ratio at low metallicity derived from the
observations is about $4\,10^{-9}$ (Parizot \& Drury 2000b).  This can
be achieve either by the CRS spectrum model, provided $x \sim 50\%$,
or by the SB spectrum model, provided $x \sim 2 or 3\%$ (see Fig.~1). 
So far, both models are equally acceptable since we chose not to
accept any prejudice about the value of $x$ from outside the
restricted field of LiBeB evolution.  But when considering the
transition metallicity associated with the two possible models, we see
that the CRS spectrum implies $Z_{\mathrm{t}}\ga 10^{-1}Z_{\odot}$,
well outside the range derived by Fields et al.  On the other hand,
the value of $Z_{\mathrm{t}}$ predicted by SB spectrum model falls
exactly in the required range.  As a conclusion, the SB model is fully
consistent with the observations provided that i) the SBEP spectrum is
flattened at low energy (in $E^{-1}$), and ii) the SN ejecta amount to
a few percent of all the matter present inside SBs.

Now let us extend the scope of our study.  Quite remarkably, the first
condition above is exactly what is expected from the SB acceleration
model developed by Bykov et al.  As for the second condition, it is in
perfect agreement with the dynamical model for SB evolution worked out
by Mac Low \& Mc Cray (1988).  In other words, had we looked
beforehand for a theoretically preferred value of $x$, we would have
chosen just the particular value which turns out to account for the
various constraints of Be Galactic evolution.  Therefore, our results
actually bring support not only to the SB model as the natural
framework for Be and B evolution studies, but also to the SB
acceleration model and standard SB dynamics.

\begin{figure}
\plottwo{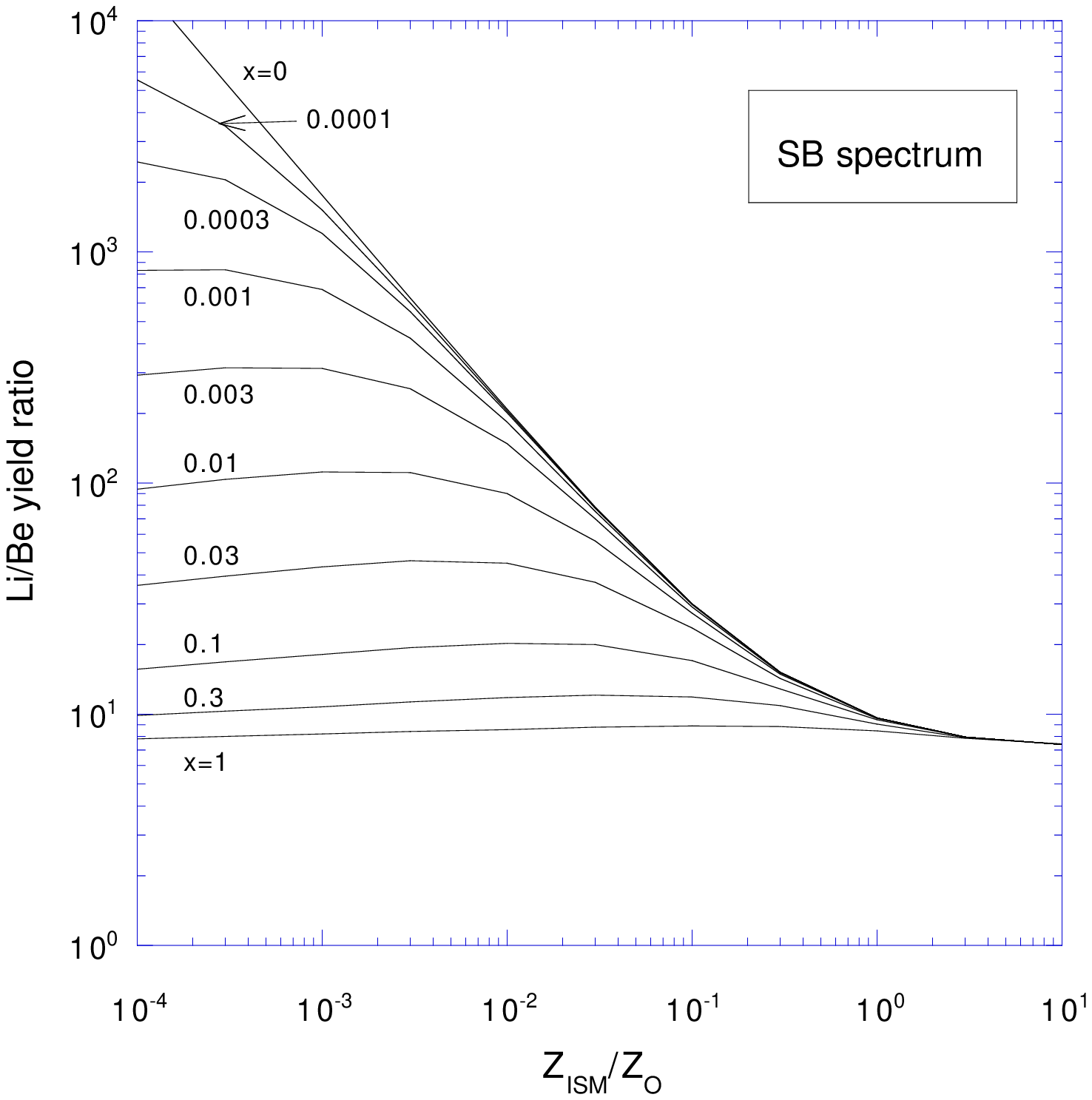}{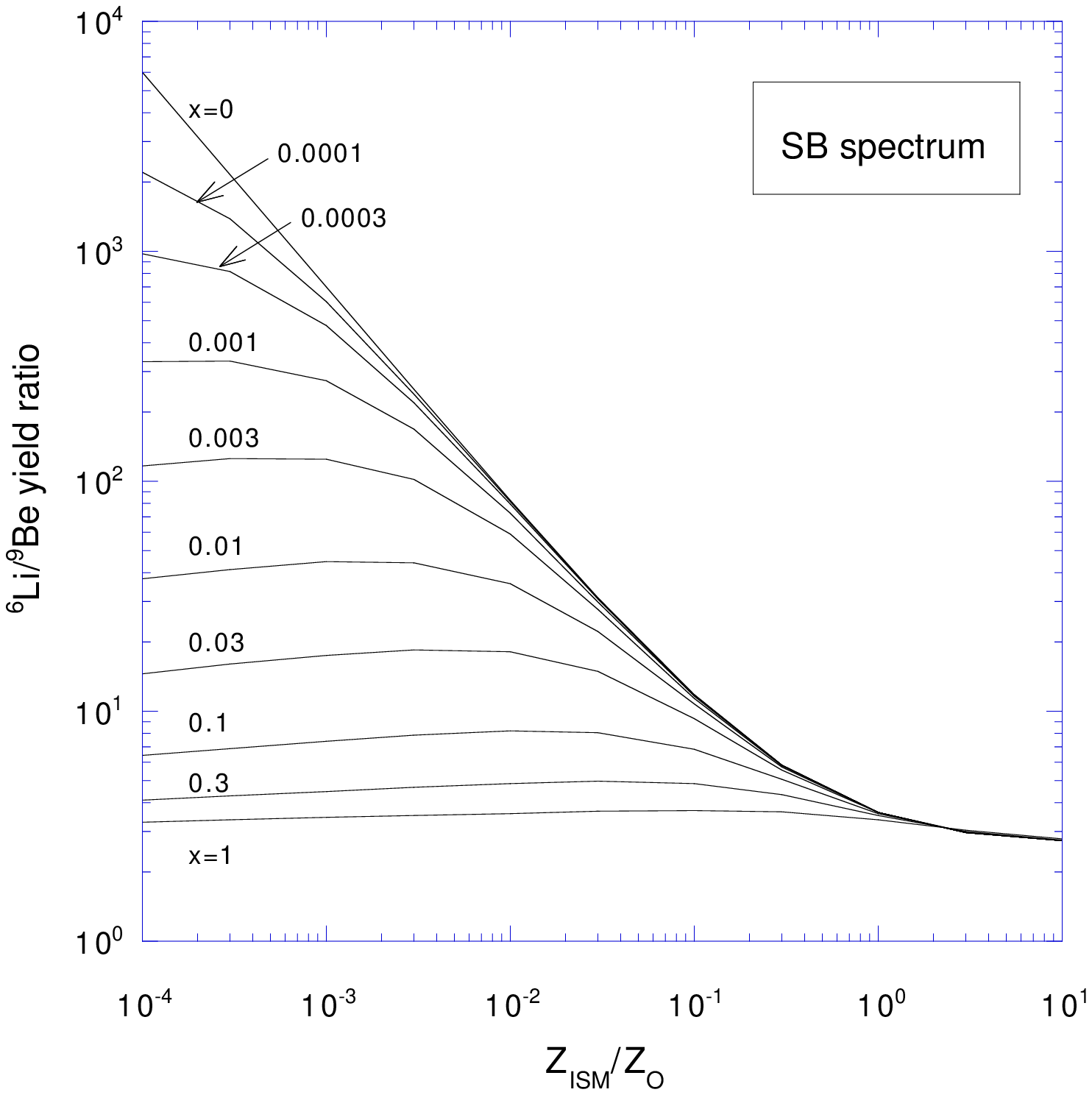}
\caption{Li/Be (left) and $^{6}$Li/$^{9}$Be (right) production ratios
obtained with the SB spectrum model, as a function of the ambient
metallicity, for various values of the mixing parameter, $x$.}
\end{figure}

Concerning B, unfortunately, only qualitative constraints can be
checked against the SB model (successfully in this instance), since
either a significant $\nu$-process or a LECR component is required
anyway to account for the observed B/Be and
$^{11}\mathrm{B}/^{10}\mathrm{B}$ ratios.  However, Li does provide
additional quantitative constraints.  First, in order not to break the
Spite plateau, the Li/Be production ratio must be lower than $\sim
100$.  This is shown to be satisfied for any value of $x$ greater than
about 1\% in Fig.~(2a).  Second, the measurement of the $^{6}$Li
abundance in two halo stars of metallicity $Z \simeq
10^{-2.3}Z_{\odot}$ indicates that the $^{6}$Li/$^{9}$Be ratio in
these stars should be in the range 20--80 (see Vangioni-Flam, Cass\'e,
\& Audouze 2000 and references therein), in contrast with the solar
value of $\sim 6$.  This could not be explained if the proportion of
SN ejecta inside SBs were of the order of 50\% (CRS spectrum model). 
However, it is quite remarkable again that the value of a few percent
derived from the SB spectrum model is totally consistent with the
observed value of the $^{6}$Li/$^{9}$Be ratio, both a low metallicity
and at solar metallicity.

\section{Conclusion}

The SB model described above has been shown to be fully consistent
with the qualitative and quantitative constraints of LiBeB Galactic
evolution: 1) it explains the inferred two-slope behaviour in the
framework of one sole model; 2) it provides the correct value of Be/O
at low metallicity; 3) it predicts the correct value of the transition
metallicity; 4) it does not break the Spite plateau; 5) it is
consistent with the $^{6}$Li/$^{9}$Be ratio at any metallicity.  Most
importantly, these successes rely on the value of only one free
parameter, namely the proportion of SN ejecta inside a SB. The value
which we find is of the order of a few percent, i.e. exactly in the
range derived from standard SB dynamical evolution.  Likewise, the SB
model is found to be successful only if the SBEPs have the SB
spectrum, i.e a flattened shape at low energy (in $E^{-1}$).  But this
is exactly what is predicted by the SB acceleration model of Bykov et
al.  As a conclusion, the SB model appears to account for all the
available constraints about LiBeB evolution by making only the most
standard assumptions about the involved models relating to other
fields of astrophysics.  This may be considered as lending support to
these models as well.

\acknowledgments This work were supported by the TMR programme of the European
Union under contract FMRX-CT98-0168.

\end{document}